\documentclass[aps,prd,showpacs,floatfix,preprint,a4paper]{revtex4-1}
\usepackage{amsfonts,amsmath,units,wasysym,epsfig,graphicx,verbatim,color,subfigure,graphicx,bm,mathrsfs,lipsum,hyperref}
\usepackage[utf8]{inputenc}
\usepackage[normalem]{ulem}  % \sout{old text} for strikeout

\begin{document}
\newcommand{\IUCAA}{Inter-University Centre for Astronomy and
  Astrophysics, Post Bag 4, Ganeshkhind, Pune 411 007, India}
\title{Relativistic tidal properties of superfluid neutron stars}
\author{Prasanta Char}\email{pchar@iucaa.in}
\author{Sayak Datta}\email{skdatta@iucaa.in} 
\affiliation{\IUCAA}
\date{\today}

\begin{abstract}
We investigate the tidal deformability of a superfluid neutron star. We calculate the equilibrium structure in the general relativistic two-fluid formalism with entrainment effect where we take neutron superfluid as one fluid and the other fluid is comprised of protons and electrons, making it a charge neutral fluid. We use a relativistic mean field model for the equation of state of matter where the interaction between baryons is mediated by the exchange $\sigma$, $\omega$ and $\rho$ mesons. Then, we study the linear, static $l=2$ perturbation on the star to compute the electric-type Love number following Hinderer's prescription. 
\end{abstract}

\maketitle

\section{Introduction}
After the observation of gravitational wave (GW) from the binary neutron star (BNS) merger event GW170817 \cite{Abbott2017}, the new era of GW astronomy has begun. Future detections of these type of events will provide us a new tool to probe the state and the composition of matter at supranuclear densities that exists inside the neutron star (NS). We can get independent information about the internal structure of the NS arising from each stage of the binary evolution \cite{Agathos2015,Takami2014,Bose2018}. It has been shown that in the inspiral stage, the GW waveform deviates significantly from the point-particle structure due to the tidal distortion in the component NS induced by the binary companion. As this deviation depends strongly on the equation of state (EOS) of the matter, it can be used to constrain the properties of matter at extreme condition. Flanagan and Hinderer have shown that at early inspiral, the correction in the waveform phase can be described by a single EOS-dependent quantity $\lambda$, known as the tidal deformability \cite{Flanagan2008,Hinderer2008,Binnington2009,Damour2009,Hinderer2010}. The quantity $\lambda$ is related to the NS radius via the relation $\lambda=\frac{2}{3}k_2R^5$, where $k_2$ is called the Love number. Therefore, it is not surprising that a huge effort has been made in the recent years to understand how $k_2$ and $\lambda$ depend on the EOS of matter, how they modify the inspiral waveform and what is the detectability of such signals to distinguish between the EOS \cite{Vines2011,Damour2012,Read2013,DelPozzo2013,Wade2014,Favata2014,Hotokezaka2016}. 

However, a very important feature of the NS matter has been mostly overlooked in those studies: superfluidity of nucleons \cite{Migdal1959,Clark1992}. There exists both theoretical and observational reasons to believe that the NS core should contain superfluid neutrons and superconducting protons. As the temperature inside the star goes below the critical transition temperature, the attractive component of the strong nuclear force leads to the BCS-like pair formation \cite{Sedrakian2018}. From the observational point of view, pulsar glitches cannot be explained without invoking the crustal superfluidity of neutrons \cite{Baym1975,Anderson1975}. Rapid cooling of the NS in Cassiopeia A is another example which can only be explained if we consider $^3P_2$ neutron pairing inside the core \cite{Page2011,Shternin2011}. To study the dynamics of superfluidity of nucleons inside NS within the general relativistic framework is a nontrivial task. As a first approximation, several authors have attempted to include superfluidity using a two-fluid model where neutrons are taken as superfluid and all other matter (including protons and electrons) are taken together as a single normal fluid \cite{Langois1998,Carter1995,Comer1994,Carter1989,carter1998 1,carter1998 2,Prix2000}. Then, the whole scenario is complicated by the ``entrainment" effect between the two fluids, where momentum of one carries some mass current of the other \cite{Sauls1989}. It has been found that the superfluidity modifies the rotational equilibrium structure \cite{Andersson2001,Prix2002}. It also affects the modes of oscillation of the star due to the generation of purely superfluid modes which are otherwise absent in the single fluid scenario \cite{Comer1999,Comer2002,Andersson2002}. Therefore, one can expect that there would be some definite signature of superfluidity in the GW waveform also. Hence, to model the tidal interaction properly, one should include superfluidity in the calculation. This problem has been investigated by Yu and Weinberg in Newtonian context \cite{Yu2017_1,Yu2017_2}. But, to our knowledge, there exists no relativistic analog for such calculation of $k_2$. So the goal of the present work is to find a Hinderer-type solution for tidal perturbation including superfluidity. For this purpose, the local properties of matter are determined using a realistic EOS model. Comer and Joynt have calculated the entrainment for the first time in a Walecka-type $\sigma$-$\omega$ model with relativistic mean field (RMF) approximation \cite{Comer2003,Comer2004}. As the matter inside the NS is highly asymmetric, the model is recently updated to include the effect of isospin dependence by Kheto and Bandyopadhyay \cite{Kheto2014,Kheto2015}. We use their updated relativistic $\sigma$-$\omega$-$\rho$ model to calculate the  ``master function" of the two-fluid formalism and the other matter variables.

The paper is organized as follows. In Sec. \ref{method}, we first discuss the two-fluid formalism followed by the calculation of the equilibrium structure along with a brief overview the RMF model of dense matter to calculate the assorted matter coefficients of the model.  Next, in Sec. \ref{tidal}, we derive the framework for tidal perturbation in the two-fluid model. Then, in Sec. \ref{res} we discuss our results. Finally, we summarize in Sec. \ref{summa}. We assume $c=G=1$ and use the metric signature $\{-,+,+,+\}$ throughout the article.

\section{General Relativistic Superfluid Neutron Star} \label{method}
In this section, we discuss the key ingredients of the superfluid formalism developed by different groups \cite{Carter1989,Comer1994,Carter1995,carter1998 1,carter1998 2,Langois1998,Comer1999,Prix2000,Andersson2001}. We adopt a two-fluid model with entrainment. The central quantity of this formalism is the master function $\Lambda$. It is a function of three scalars, $n^2 = -n^{\mu}n_{\mu}$, $p^2 = -p^{\mu}p_{\mu}$, and $x^2 = -n^{\mu}p_{\mu}$. These scalars are constructed from the conserved number density currents $n^{\mu}$ and $p^{\mu}$ of the neutron and proton, respectively. When the fluids are comoving, the total thermodynamic energy density is $-\Lambda (n^2,p^2,x^2)$. When $\Lambda$ is given, the stress-energy tensor can be written as,
\begin{equation}
T^{\mu}_{\nu} = \Psi \delta^{\mu}_{\nu} + p^{\mu}\chi_{\nu} + n^{\mu}\mu_{\nu},
\end{equation}
where, $\Psi$ is the generalized pressure, and can be expressed as,
\begin{equation}
\label{generalised pressure}
\Psi = \Lambda - n^{\rho}\mu_{\rho}-p^{\rho}\chi_{\rho}.
\end{equation}
Here, $\chi_{\nu}$ and $\mu_{\nu}$ are the chemical potential covectors of the proton and the neutron fluid respectively.
\begin{equation}
\label{momentum covector eqn}
\mu_{\mu} = {\cal B} n_{\mu} + {\cal A} p_{\mu}, ~~~~ \chi_{\mu} = {\cal C} p_{\mu} + {\cal A} n_{\mu},
\end{equation}
and, the ${\cal A, B}$ and ${\cal C}$ coefficients are given by,
\begin{equation}
{\cal A} = -\frac{\partial\Lambda}{\partial x^2},~~~ {\cal B} = -2\frac{\partial\Lambda}{\partial n^2},~~~ {\cal C} = -2\frac{\partial\Lambda}{\partial p^2}.
\end{equation}
The quantities $\mu_{\mu}$ and $\chi_{\mu}$ make the so-called entrainment effect vivid, that can be understood from the Eq. \ref{momentum covector eqn}. The momentum of one component carries along some of the mass current of the other component as long as ${\cal A} \neq 0$. As a result the master function becomes ``entrainment-free" if ${\cal A} = 0$, implying that it does not depend on $x^2$. 
The equations of motion also consist of two conservation equations for $n^{\mu}$ and $p^{\mu}$,
\begin{equation}
\label{conservation of number}
\nabla_{\mu}n^{\mu} = \nabla_{\mu}p^{\mu} = 0.
\end{equation}
We also have a set of Euler type equations \cite{Comer1999},
\begin{equation}
n^{\mu}\nabla{_{[\mu}\mu_{\nu]}} = p^{\mu}\nabla{_{[\mu}\chi_{\nu]}} = 0,
\end{equation}
where, the square braces represent the antisymmetrization of the closed indices.

\subsection{Equation of state of neutron star matter} \label{matter}
We calculate the master function using a $\sigma$-$\omega$-$\rho$ model with self-interaction in the RMF approximation. The Lagrangian of the theory is given by,
\begin{eqnarray}\label{lag}
{\cal L}_B &=& \sum_{B=n,p} \bar\Psi_{B}\left(i\gamma_\mu{\partial^\mu} - m_B
+ g_{\sigma B} \sigma - g_{\omega B} \gamma_\mu \omega^\mu
- g_{\rho B}
\gamma_\mu{\mbox{\boldmath $\tau$}}_B \cdot
{\mbox{\boldmath $\rho$}}^\mu \right)\Psi_B\nonumber\\
&& - \frac{1}{2} \partial_\mu \sigma\partial^\mu \sigma
-\frac{1}{2} m_\sigma^2 \sigma^2 - \frac{1}{3}b m \left(g_\sigma \sigma\right)^3 
- \frac{1}{4}c  \left(g_\sigma \sigma\right)^4 \nonumber\\
&& -\frac{1}{4} \Omega_{\mu\nu}\Omega^{\mu\nu}
-\frac{1}{2}m_\omega^2 \omega_\mu \omega^\mu
- \frac{1}{4}{\mbox {\boldmath $\mathrm{P}$}}_{\mu\nu} \cdot
{\mbox {\boldmath $\mathrm{P}$}}^{\mu\nu}
- \frac{1}{2}m_\rho^2 {\mbox {\boldmath $\rho$}}_\mu \cdot
{\mbox {\boldmath $\rho$}}^\mu ~,
\end{eqnarray}
where, $m_B$ is the baryon mass. In our calculations, we use the nucleon mass $m$ as the average of the baryon masses. The Dirac effective mass $m_*$ is defined as $m_* = m- g_\sigma \sigma$. The $\sigma$, $\omega$ and $\rho$ represent the scalar, vector and vector-isovector interactions respectively. The ${\mbox{\boldmath $\tau$}}_B$ is the isospin operator. The $\Omega_{\mu \nu}$, ${\mbox{\boldmath $\mathrm{P}$}}_{\mu \nu}$ are the field tensors for $\omega$ and $\rho$ mesons respectively. For the two-fluid system, a frame is chosen in such a way that neutrons have zero spatial momentum and the proton momentum have a boost along the z-direction as $k_p^\mu = \left(k_0, 0, 0, K\right)$. We identically follow the same procedure as Ref. \cite{Kheto2014,Kheto2015} to solve the meson field equations and numerically evaluate the master function $\Lambda$, generalized pressure $\Psi$ etc. in the limit $K \rightarrow 0$.

\subsection{Equilibrium structure} \label{equil}
The background metric of the star under consideration is taken to be static and spherically symmetric. Therefore, it is possible to write the metric in the Schwarzschild form as follows,

\begin{equation}
\label{equilibrium metric}
ds_0^2 = g^{(0)}_{\alpha\beta} dx^{\alpha} dx^{\beta}= -e^{\nu(r)} dt^2 + e^{\kappa(r)} dr^2 + r^2 (d\theta^2 + \sin^2\theta d\phi^2)
\end{equation}
We evaluate the two metric functions from the Einstein's equations as,
\begin{eqnarray}
\kappa^{\prime} &=& \frac{1-e^{\kappa}}{r} - 8\pi r e^{\kappa}\Lambda |_0, \nonumber\\
\nu^{\prime} &=& -\frac{1-e^{\kappa}}{r}  + 8\pi r e^{\kappa}\Psi |_0,
\label{tov_munu}
\end{eqnarray}
The radial profiles for $n(r)$ and $p(r)$ are determined by the following equations \cite{Comer1999},
\begin{eqnarray}
{\cal A}^0_0 |_0 p' + {\cal B}^0_0 |_0 n' + \frac{1}{2} \mu|_0\nu ' &=& 0, \nonumber\\
{\cal C}^0_0 |_0 p' + {\cal A}^0_0 |_0 n' + \frac{1}{2} \chi|_0\nu ' &=& 0,
\label{tov_np}
\end{eqnarray}
where, 
\begin{equation}
\begin{split}
{\cal A}^0_0 &= {\cal A} +2\frac{\partial {\cal B}}{\partial p^2} np + 2\frac{\partial {\cal A}}{\partial n^2}n^2 + 2\frac{\partial {\cal A}}{\partial p^2}p^2 + \frac{\partial {\cal A}}{\partial x^2} np,\\
{\cal B}^0_0 &= {\cal B} +2\frac{\partial {\cal B}}{\partial n^2}n^2 + 4\frac{\partial {\cal A}}{\partial n^2}np + \frac{\partial {\cal A}}{\partial x^2}p^2,\\
{\cal C}^0_0 &= {\cal C} +2\frac{\partial {\cal C}}{\partial p^2}p^2 + 4\frac{\partial {\cal A}}{\partial p^2}np + \frac{\partial {\cal A}}{\partial x^2}n^2.
\end{split}
\end{equation}
The variables that are more appropriate for the RMF calculations are the two Fermi wave numbers $k_n$ and $k_p$. Hence, we substitute everywhere the number densities with the Fermi wave numbers by using $n = \frac{k_n^3}{3\pi^2}$ and $p = \frac{k_p^3}{3\pi^2}$, and solve for $k_n$ and $k_p$ instead. A convenient way for determining the Dirac effective mass $m_*|_0(k_n,k_p)$ has been discussed in \cite{Comer2003}. They have turned the transcendental algebraic relation in Eq. \ref{m star eqn} into a differential equation via
\begin{equation}
m_*' |_0 = \frac{\partial m_*}{\partial k_n}\bigg |_0 k_n' + \frac{\partial m_*}{\partial k_p}\bigg |_0 k_p',
\end{equation}
where $k_n'$ and $k_p'$ are obtained from Eq. \ref{tov_np}.
The prime in the equations represent a radial derivative and a zero subscript represents that after the partial derivatives are taken, we take $K\rightarrow 0$.
The boundary conditions are put at the center and the surface of the star. Non-singularity at center imposes $\kappa(0) = 0$ and $\kappa'(0)$ and $\nu'(0)$ vanish too. This condition along with Eq. \ref{tov_np} imposes that $k_n'(0) = k_p'(0) = 0$. The continuity of the metric variable at the surface of the star $r=R$, implies that the total mass of the star is,
\begin{equation}
M = -4\pi \int^R_0 dr r^2 \Lambda|_0(r),
\label{tov_mass}
\end{equation}
and $\Psi|_0(R) = 0$. We have written down the necessary expressions for all the matter quantities used in our calculations $\left(\Lambda|_0, \Psi|_0, \mu|_0, \chi|_0, m_*|_0 , {\cal A}|_0, {\cal B}|_0, {\cal C}|_0, {\cal A}^0_0|_0, {\cal B}^0_0|_0, {\cal C}^0_0|_0, \left.\frac{\partial m_*}{\partial k_n}\right|_0, \left.\frac{\partial m_*}{\partial k_p}\right|_0 \right)$  in the Appendix.

%\subsection{Even Parity Static Perturbation Equations}
\section{Perturbed Superfluid Star}\label{tidal}
When a nonrotating static NS is placed in an external time-independent tidal field ${\cal E}_{ij}$, the induced quadrupolar response can be written as,
 \begin{equation}
  Q_{ij} = - \lambda \mathcal{E}_{ij},
 \end{equation}
where, $\lambda$ is the tidal deformability that is related to $l=2$ dimensionless Love number $k_2$ by $k_2=\frac{3}{2}\lambda R^{-5}$. 
To calculate $k_2$, we consider the linearized perturbations on the static and spherically symmetric star following Thorne and Campolattaro \cite{Thorne1967}. Then, we can write the full metric as,
\begin{equation}
g_{\alpha\beta} = g^{(0)}_{\alpha\beta} + h_{\alpha\beta},
\end{equation}
where, $g^{(0)}_{\alpha\beta}$ and $h_{\alpha\beta}$ are the background and the perturbed part of the metric respectively. We expand the metric and the fluid perturbations in terms of the spherical harmonics $Y^m_l(\theta, \phi)$, keeping only the  $m=0$ terms due to spherical symmetry of the background \cite{Chandra}. We focus only on the $l=2$, static, even-parity perturbations in the Regge-Wheeler gauge \cite{Regge1957}. Replacing $Y^0_l(\theta, \phi)=P_l(\theta)$, it is written as,
\begin{equation}
\label{perturbed metric}
\begin{split}
&h_{\alpha\beta} = {\rm diag}[-e^{\nu(r)}H_{0}(r), e^{\kappa(r)}H_2(r), r^2K(r), r^2\sin^2\theta K(r)] P_2(\theta).
\end{split}
\end{equation}
It is straightforward to calculate the perturbation in the energy momentum tensor. We have $\delta T^0_0 = \delta \Lambda$ and $\delta T^i_j = \delta \Psi\delta^i_j$. We use these relations in the linearized Einstein equations $\delta G^{\alpha}_{\beta} = 8\pi \delta T^{\alpha}_{\beta}$ and find the equations governing the metric perturbations. From $\delta G^{\theta}_{\theta}-\delta G^{\phi}_{\phi} = 0$ and $\delta G^{r}_{\theta} = 0$, it follows that   $H_0 = -H_2 \equiv H$ and $K^{\prime} + H^{\prime} + H \nu^{\prime} = 0$ respectively. After using $\delta G^{\theta}_{\theta}+\delta G^{\phi}_{\phi} = 16 \pi \delta \Psi$, we find $2\delta \Psi = P_2(\theta) H ( \Lambda-\Psi)$. Finally $\delta  G^r_r = 8\pi \delta T^r_r$ relates $K, H$ and $H'$ as follows,
%Using these perturbation in Einstein equation and linearizing them we can find the perturbed metric equation as follows,
\begin{equation}
\begin{split} K = \frac{-r^2 \nu^{\prime} H^{\prime}}{4e^{\kappa}} + \frac{H\{2-r^2\nu^{\prime 2} + e^{\kappa} (8\pi r^2 (\Psi - \Lambda)-6)\}}{4e^{\kappa}}
\end{split}
\end{equation}
From the expressions of $\delta \mu_{\rho}$ and $\delta \chi_{\rho}$ found by Comer \cite{Comer1999}, it follows,
\begin{equation}
\label{mu-chi equation}
\begin{split}
&\delta\mu_0 = ({\cal A}^0_0\delta p + {\cal B}^0_0 \delta n)u^0 
g_{00}+u^0\frac{\mu}{2} h_{00}\\
&\delta\chi_0 = ({\cal A}^0_0\delta n + {\cal C}^0_0 \delta p)u^0 
g_{00}+u^0\frac{\chi}{2} h_{00}.
\end{split}
\end{equation}
From the linearized Euler equation, it can be found that \cite{Comer1999},
\begin{equation}
\label{perturbed Euler}
\partial_t\delta\mu_i = \partial_i\delta\mu_t,\,\,\,\,\partial_t\delta\chi_i = \partial_i\delta\chi_t.
\end{equation}
Therefore, staticity implies $\delta\mu_0 = \delta\chi_0 = 0$. Using Eq. \ref{perturbed metric}, \ref{mu-chi equation} and \ref{perturbed Euler},  we find,
\begin{equation}
\begin{split}
&\delta n = \frac{(\chi {\cal A}^0_0-\mu {\cal C}^0_0)}{({\cal B}^0_0{\cal C}^0_0-{\cal A}^{02}_0)}\frac{HP_2(\theta)}{2}\\
&\delta p = \frac{(\mu {\cal A}^0_0-\chi {\cal B}^0_0)}{({\cal B}^0_0{\cal C}^0_0-{\cal A}^{02}_0)}\frac{HP_2(\theta)}{2}
\end{split}
\end{equation}
Next we calculate $\delta \Lambda$ as,
\begin{equation}
\begin{split}
\delta \Lambda &= \frac{\partial \Lambda}{\partial x^2}\delta x^2+\frac{\partial \Lambda}{\partial p^2}\delta p^2+\frac{\partial \Lambda}{\partial n^2}\delta n^2\\
&=-[({\cal A}n+{\cal C}p)\delta p+({\cal A}p+{\cal B}n)\delta n]\\
&= -g\frac{H}{2}P_2(\theta),
\end{split}
\end{equation}
where,
 \begin{equation}
\begin{split}
g = \frac{\mu^2 {\cal C}^0_0+\chi^2 {\cal B}^0_0-2\mu\chi {\cal A}^0_0}{ {\cal A}^{02}_0-{\cal B}^0_0 {\cal C}^0_0}.
\end{split}
\end{equation}
To get the final perturbation equation, we use the following Einstein equation along with the expression of $\delta \Lambda$,
\begin{equation}
\label{final Einstein equation}
\delta G^{t}_{t} - \delta G^{r}_{r}= -4\pi g H P_2(\theta)+4P_2(\theta)\pi  H (\Psi - \Lambda).
\end{equation}
After some simplifications, it reduces to,
\begin{widetext}
\begin{equation}
\label{H equation}
\begin{split}
H''+& H' \big[4 \pi  re^{\kappa } (\Lambda +\Psi) +\frac{e^{\kappa}+1}{r}\big]+H[ 4\pi e^{\kappa }   (-g+9 \Psi -5 \Lambda)-\nu '^2-\frac{6 e^{\kappa }}{r^2}]=0.
\end{split}
\end{equation}
\end{widetext}
This is the central equation for the determination of the tidal Love numbers. It is also to be noted here that Eq. \ref{H equation} contains the coefficients ${\cal A}_{\mu\nu}$, ${\cal B}_{\mu\nu}$ and ${\cal C}_{\mu\nu}$ which we evaluated for the equilibrium configuration. We now emphasize the main difference between Eq. \ref{H equation} and its nonsuperfluid single fluid counterpart Eq. 15 of Hinderer \cite{Hinderer2008}. In case of the normal fluid, barotropic nature of the fluid is assumed. Hence, it is possible to write $\delta\rho = \frac{d\rho}{dp}\delta p$ and substitute it into one of the perturbed Einstein's equations. But, for any multifluid scenarios, this assumption is not true, in general. Therefore, we explicitly calculate the $\delta\Lambda$ with respect to the fluid variables and the perturbed metric variables. Thus, the final equation gets modified and so does the response to the perturbation subsequently.

\subsection*{Calculating the tidal Love number} 
In order to calculate the tidal deformability, one needs to solve Eq. \ref{H equation} numerically inside the NS and match it with the external solution of the same equation on the surface of the star. This has been discussed extensively in \cite{Hinderer2008,Binnington2009,Damour2009}.   We discuss only the initial conditions here. We need to integrate Eq. \ref{H equation} for metric perturbation function $H$ radially outward from the center using the profiles of the background quantities calculated from TOV equations. But for numerical purposes, we can not start from $r=0$, rather we use a very small cutoff radius $(r = r_0 = 10^{-6})$. The boundary condition for Eq. \ref{H equation}  around the regular singular point $r=0$ can be taken to be $H(r) \sim \bar{h}r^2$, with $\bar{h}$ an arbitrary constant. As this equation is homogeneous in $H$ and the tidal deformability depends explicitly on the value of $y ~(= \frac{r H^{\prime}}{H})$ at the surface, the scaling constant $\bar{h}$ is irrelevant. So, the starting value for the metric variable can be chosen as, $H(r_0) = r_0^2$ and $H'(r_0) = 2r_0$.

The value of the deformability can be computed with respect to $y$ and the compactness $C= \frac{M}{R}$, by matching the internal and external value of $H$ at surface. The tidal Love number $k_2$ then takes the functional form \cite{Hinderer2008,Binnington2009,Damour2009},
\begin{widetext}
\begin{equation}
\label{expr_k2}
\begin{split}
k_2 &= \frac{8}{5}(1-2C)^2C^5\big[2C(y-1)-y+2\big]\bigg[2C(4(y+1)C^4 + (6y-4)C^3+(26-22y)C^2\\
&+3(5y-8)C-3y+6)-3(1-2C)^2(2C(y-1)-y+2)\log(\frac{1}{1-2C})\bigg]^{-1}.
\end{split}
\end{equation}
\end{widetext}
This expression of $k_2$ is similar to the one fluid formalism. This happens because the information of the fluid enters through $y|_{r=R}$ and $C$. Two-fluid model does not change the external solution. It only changes the internal equation of $H$, that gives us different value of $y|_{r=R}$, leading to the change in the value of  $k_2$ but not its expression.

\section{Results and Discussions}\label{res}
Now, we discuss the numerical results for tidally deformed superfluid stars. First, we calculate the static equilibrium configurations by solving the TOV equations using realistic EOS. As only a few calculations are available for the two-fluid system, we choose a RMF model with scalar self-interaction terms and use NL3 and GM1 parametrizations. In this context, we impose the $\beta$-equilibrium at the center of the star by imposing the condition, $\mu|_0 = \chi|_0$ to get a set of $k_n$, $k_p$ and $m_*$ for calculating the central number densities of neutron and proton, energy density ($-\Lambda|_0$) and pressure ($\Psi|_0$). These quantities serve as the starting point for solving Equations \ref{tov_munu}, \ref{tov_np} and \ref{tov_mass} to find the structure of the star and generate profiles for various background quantities for several different sets of ($k_n, k_p, m_* $) corresponding to different central energy densities. The maximum mass, we have calculated for NL3 is $2.793~\textnormal{M}_\odot$ and the corresponding radius being 13.19 km. Similarly, for GM1, the maximum mass is found to be $2.384~\textnormal{M}_\odot$ and the corresponding radius is 11.9 km. Details of those parameter sets are listed in Table \ref{tab1}. Here, we also stress the fact that these EOS serve representative purposes only as both of them are ruled out by the latest GW data.

\begin{table} 
\caption{Nucleon-meson coupling constants in 
the NL3 and GM1 sets are taken from Refs.\cite{Glendenning1991,Fattoyev2010}. The coupling constants are obtained by reproducing the saturation properties of symmetric nuclear matter as detailed in the text. All the parameters are in $fm^{2}$, except $b$ and $c$ which are dimensionless.}

\begin{center}
%\vspace{5cm}
\begin{tabular}{cccccc} 

\hline\hline
\hfil& 
$c_{\sigma}^2$& $c_{\omega}^2$& $c_{\rho}^2$& $b$& $c$ \\ \hline
NL3& 15.739& 10.530& 5.324& 0.002055& -0.002650 \\ \hline
GM1& 11.785&  7.148& 4.410& 0.002948& -0.001071 \\ \hline
\hline

\end{tabular}
\end{center}
\label{tab1}
\end{table}

Then, we solve the metric perturbation Eq. \ref{H equation} using the background profiles mentioned earlier, find $y$ at the surface of the stars and calculate the Love numbers using Eq. \ref{expr_k2}. The effect of entrainment is implicitly manifested in the results as it enters the equation for $H$ via the function $g$ which contains the ${\cal A}$ coefficient. In Table \ref{tab2}, we have compared our calculated values for Love numbers and dimensionless tidal deformabilities (defined as, $\Lambda_T = \lambda/M^5$)  for single fluid and two fluid stars for the NL3 EOS in the mass range of $1\textnormal{M}_\odot$ to $2\textnormal{M}_\odot$. Similar results are shown in Tab. \ref{tab3} for the GM1 EOS. The change in tidal deformability between the two cases is $\Delta\Lambda_T =  \Lambda_T^{2\textnormal{-fluid}} - \Lambda_T^{1\textnormal{-fluid}}$. We show the percentage change $(\Delta\Lambda_T/\Lambda_T^{1\textnormal{-fluid}})$ in the tables for the two chosen EOS. For all the stellar configurations, we find the $\Lambda_T^{2\textnormal{-fluid}}$ is larger than $\Lambda_T^{1\textnormal{-fluid}}$. The change is about $\sim 5\%-8 \%$ for NL3 EOS and and $\sim 4\%-11 \%$ for GM1 EOS. We find a trend of increase in the percentage change as we go to higher mass.

\begin{table}
\caption{Comparison of Love numbers calculated using both one fluid and two fluid approach for NL3 parameter set}
\begin{tabular}{|c|c|c|c|c|c|} 
\hline
Mass (M$_\odot$)& $k_2^{1\textnormal{-fluid}}$ & $k_2^{2\textnormal{-fluid}}$  & $\Lambda_T^{1\textnormal{-fluid}}$ & $\Lambda_T^{2\textnormal{-fluid}}$ &$\Delta \Lambda_T/\Lambda_T^{1\textnormal{-fluid}}$(\%)\\
\hline
1.0 & 0.1288 & 0.1877 & 7814.7 & 8289.8 & 6.07\\
1.1 & 0.1249 & 0.175 & 4746.8 & 4997.8 & 5.29\\
1.2 & 0.1201 & 0.1636 & 2981.5 & 3205.6 & 7.52\\
1.3 & 0.1148 & 0.1523 & 1928.2 & 2073.7 & 7.55\\
{\bf 1.4} & 0.1092 & 0.1416 & {\bf 1276.7} & {\bf 1382.9} & 8.32\\
1.5 & 0.1033 & 0.1312 & 862.7 & 932.2 & 8.06\\
1.6 & 0.0972 & 0.1213 & 591.5 & 640.6 & 8.30\\
1.7 & 0.0911 & 0.1119 & 411.2 & 446.5 & 8.58\\
1.8 & 0.0849 & 0.1028 & 288.7 & 313.3 & 8.52\\
1.9 & 0.0787 & 0.0941 & 204.3 & 221.9 & 8.61\\
2.0 & 0.0726 & 0.0856 & 145.3 & 157.5 & 8.40\\
\hline
\end{tabular}
\label{tab2}
\end{table}

\begin{table}
\caption{Comparison of Love numbers calculated using both one fluid and two fluid approach for GM1 parameter set}
\begin{tabular}{|c|c|c|c|c|c|} 
\hline
Mass (M$_\odot$)& $k_2^{1\textnormal{-fluid}}$ & $k_2^{2\textnormal{-fluid}}$  & $\Lambda_T^{1\textnormal{-fluid}}$ & $\Lambda_T^{2\textnormal{-fluid}}$ &$\Delta \Lambda_T/\Lambda_T^{1\textnormal{-fluid}}$(\%)\\
\hline
1.0 & 0.133 & 0.1874 & 5899.5 & 6141 & 4.09\\
1.1 & 0.1273 & 0.1731 & 3577.9 & 3755.1 & 4.95\\
1.2 & 0.1207 & 0.1597 & 2206.3 & 2342.9 & 6.19\\
1.3 & 0.1136 & 0.1468 & 1399.6 & 1495.4 & 6.84\\
{\bf 1.4} & 0.106 & 0.1343 & {\bf 903.9} & {\bf 971} & 7.42\\
1.5 & 0.0982 & 0.1223 & 591.3 & 639.3 & 8.11\\
1.6 & 0.0902 & 0.1107 & 390.2 & 424.7 & 8.84\\
1.7 & 0.0822 & 0.0995 & 258.9 & 282.9 & 9.26\\
1.8 & 0.0742 & 0.0887 & 171.8 & 189 & 10.01\\
1.9 & 0.0661 & 0.0784 & 113.4 & 125.9 & 11.02\\
2.0 & 0.058 & 0.0681 & 73.9 & 82.3 & 11.36\\
\hline
\end{tabular}
\label{tab3}
\end{table}
It is important to understand the significance of the present result in the context of constraining the dense matter EOS using the GW data. The same RMF model gives higher values of $\Lambda$ for superfluid cases. Hence, more EOS can be ruled out which are otherwise allowed if we do not consider superfluidity inside the NS.

\begin{figure}
\includegraphics[width=0.8\textwidth]{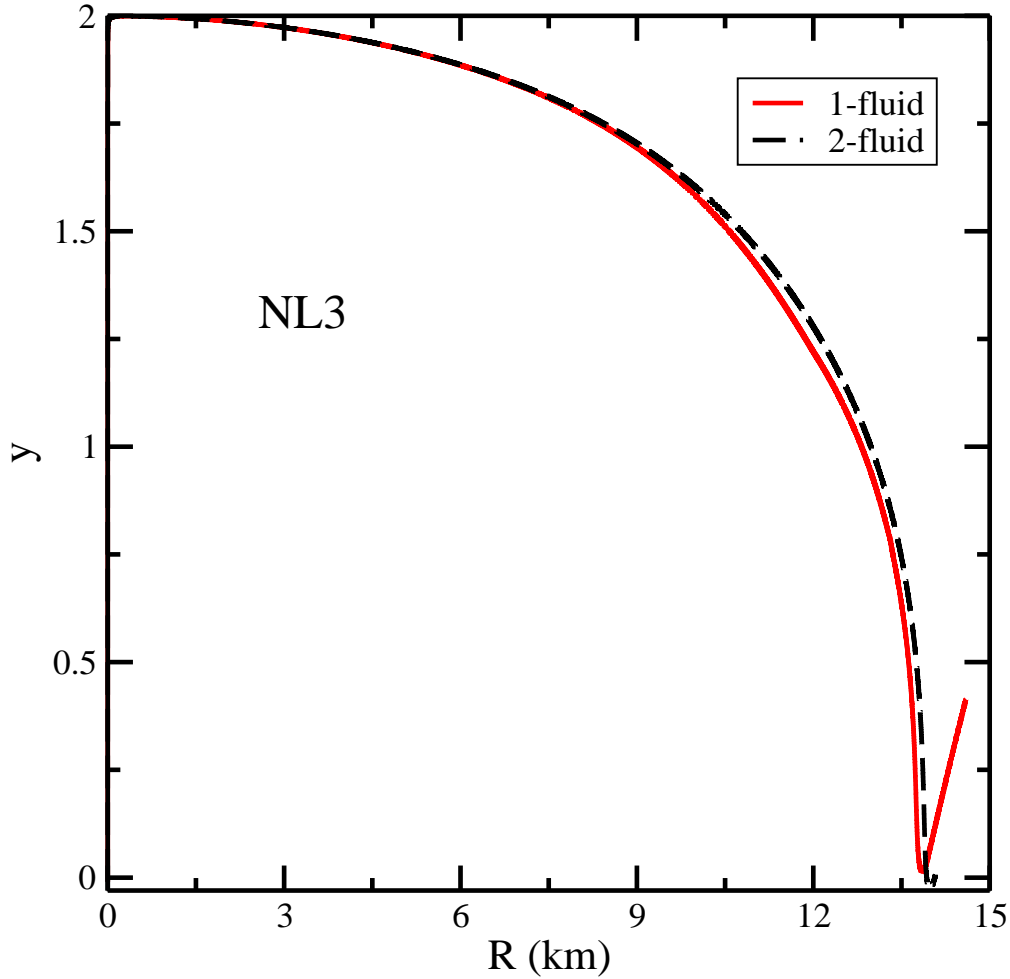}
\caption{Profile of y for NL3}
\label{fig1}
\end{figure}

\begin{figure}
\includegraphics[width=0.8\textwidth]{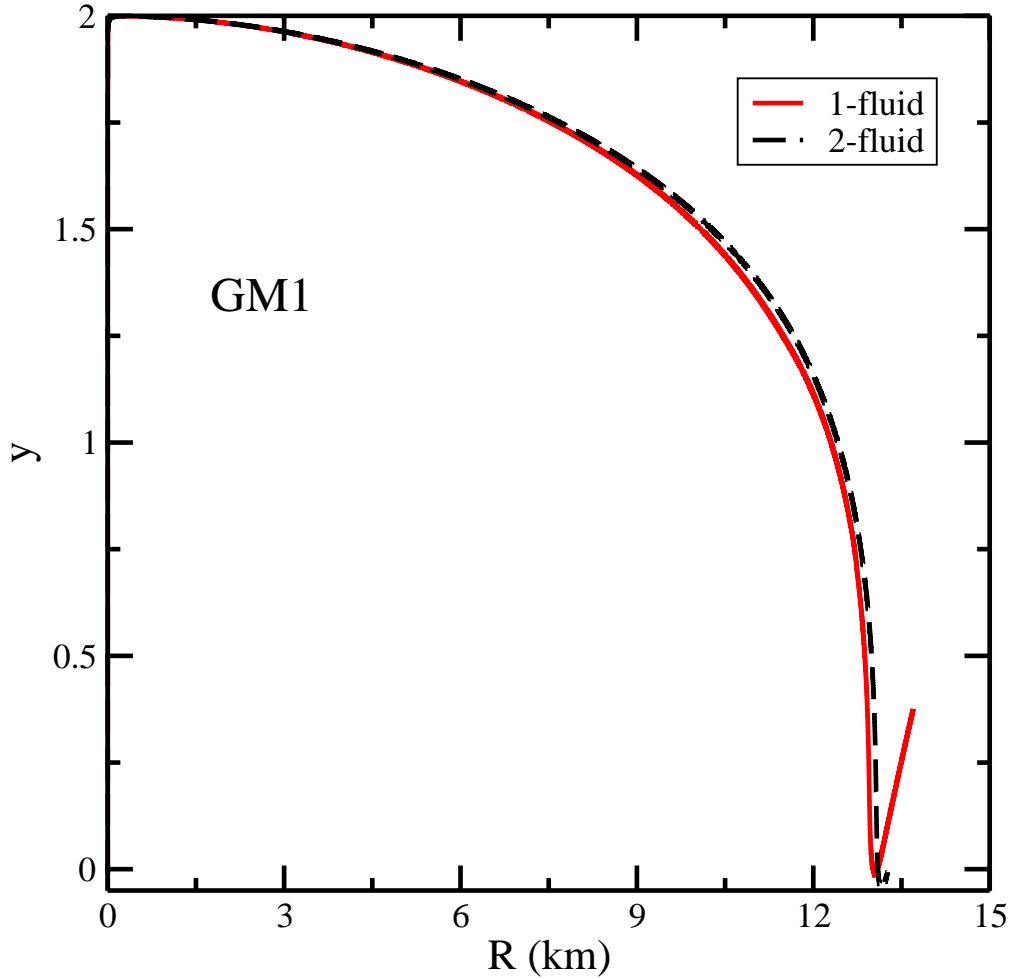}
\caption{Profile of y for GM1 {\color{red}}}
\label{fig2}
\end{figure}

We also plot the profile of $y$ of a $1.4$M$_\odot$ star for both one and two-fluid cases and compare them in Fig. \ref{fig1} and \ref{fig2} for NL3 and GM1 models respectively. We find that $y$ differs significantly near the surface of the star in both cases. This leads to different values of Love numbers for single and two-fluid cases. We find the Love numbers are usually larger for two-fluid system. To understand the reason for this particular behavior, we take resort to the perturbation analysis of two-fluid star done by Comer et al. \cite{Comer1999}. In their studies, they found the existence of several superfluid oscillation mode which cannot be found otherwise in a single fluid star. This behavior is specific to the two-fluid formalism where different fluid mode appears due to the two different types of fluid displacement. Now, Flanagan and Hinderer discussed that tidal deformation of a star can be considered as the sum of the deformations arising from different fluid modes excited inside the star, due to the tidal perturbation. Hence, we can argue that due to the appearance of extra fluid modes in the superfluid stars, we will get slightly larger deformations. This is exactly what we find in the Tables \ref{tab2} and \ref{tab3}.

\section{Summary}\label{summa}
We have investigated the possibility of probing the superfluid nature of matter inside the NS using the GW observations. We have calculated the $l=2$ tidal Love number for superfluid NS in a two-fluid model using realistic nuclear interactions and compared them with the ones calculated for non-superfluid single fluid stellar configurations. We have found the overall change in the dimensionless tidal deformability ranging from $4$ to $11\%$ corresponding to stars with mass range within $1-2$~M$_\odot$. The superfluid framework imposes stricter constraints on the dense matter EOS in the context of GW observation. We expect the other Love numbers, such as $k_3, j_2$ etc., also to be sensitive to superfluidity. This provides us the opportunity to improve our understanding of the superfluid nature of the dense matter with better observational data in future. 

The main criticism of this work can be the noninclusion of crust for the two-fluid star. We integrate the uniform matter EOS up to the surface. Therefore, the low density behavior of the system might affect the result for the low mass stars. In our defense, we are unable to incorporate a crust within the two-fluid master function formalism, as a self-consistent model of multi-fluid crust is yet to be developed. But, we can point out the result that one can expect, albeit in a crude way. If we implement a proper crust model, there will surely be some increase in the radius. But, the mass will not enhance significantly. Since, we are underestimating the radius while considering the star uniform, the values of $\Lambda$ are also being underestimated. Hence, inclusion of a crust model will increase the tidal deformability of a superfluid star which is already higher than the normal star. However, the higher mass stars have relatively thinner crustal region than their low mass counterparts. As a result, the effect of inclusion of a crust will be less prominent for the higher mass stars.
This formalism is one of the most consistent way to treat superfluid NS in general relativity. Therefore, one has to consider it to study the tidal phenomena in NS. This work can be seen as a proof-of-concept study in that direction. In a future work, we plan to incorporate the low density behavior in our model in a self-consistent manner. We also intend to present a general framework for the calculation of both electric- and magnetic-type Love numbers in case of a superfluid NS with entrainment.

\section*{acknowledgements}
The authors would like to thank A. Kheto, D. Bandyopadhyay, A. Basu and S. Bose for the useful discussions. The authors also acknowledge support from the Navajbai Ratan Tata Trust. Additionally, S. D. would like to thank the University Grants Commission (UGC), India, for the financial support as a senior research fellow.

\section*{Appendix: CALCULATIONS OF MATTER COEFFICIENTS}
The master function and the chemical potentials of neutron and proton fluids in the limit $K \rightarrow 0$ are given by, 
\begin{eqnarray}
    \left.\Lambda\right|_0 &=&  - \frac{c_\omega^2}{18 \pi^4}\left(k_n^3+k_p^3\right)^2-\frac{c_\rho^2}{72 \pi^4}\left(k_p^3-k_n^3\right)^2- \frac{1}{4 \pi^2} \left(k_n^3 
             \sqrt{k_n^2 + \left.m^2_*\right|_0} + 
             k_p^3 \sqrt{k_p^2 + \left.m^2_*\right|_0}
             \right)   \cr
             && \cr
             &&- \frac{1}{4} c_\sigma^{- 2} \left[\left(2 m - \left.m_*\right|_0\right) 
             \left(m - \left.m_*\right|_0\right)+\left.m_*\right|_0\left( b m c_\sigma^2\left(m-\left.m_*\right|_0\right)^2 
          + c  c_\sigma^2\left(m-\left.m_*\right|_0\right)^3\right)\right]\nonumber \\
          &&-\frac{1}{3} b m \left(m-\left.m_*\right|_0\right)^3\--\frac{1}{4} c  \left(m-\left.m_*\right|_0\right)^4- \frac{1}{8 \pi^2} \left( k_p \left[2 k_p^2 + m_e^2\right] \sqrt{k_p^2 + m^2_e}\right.\nonumber \\
             &&
             \left. - m^4_e {\rm ln}\left[
             \frac{k_p + \sqrt{k_p^2 + m^2_e}}{m_e}\right]\right) 
              \ , \\
             && \cr
    \left.\mu\right|_0 &=&-\frac{\pi^2}{k_n^2} \left.\frac{\partial \Lambda}{\partial k_n}\right|_0= \frac{c_\omega^2}{3 \pi^2}\left(k_n^3+k_p^3\right)
 - \frac{c_\rho^2}{12 \pi^2}\left(k_p^3-k_n^3\right)  + \sqrt{k_n^2 + 
             \left.m^2_*\right|_0} \ , \\ 
             && \cr
    \left.\chi\right|_0 &=&-\frac{\pi^2}{k_p^2} \left.\frac{\partial \Lambda}{\partial k_p}\right|_0= \frac{c_\omega^2}{3 \pi^2}\left(k_n^3+k_p^3\right)
+ \frac{c_\rho^2}{12 \pi^2}\left(k_p^3-k_n^3\right) + \sqrt{k_p^2 + 
             \left.m^2_*\right|_0} + \sqrt{k_p^2 + m_e^2}. \  
\end{eqnarray}             
The generalized pressure $\Psi$ is related to the master function with the following relation,
\begin{equation}
    \left.\Psi\right|_0 = \left.\Lambda\right|_0 + \frac{1}{3 \pi^2}
             \left(\left.\mu\right|_0 k_{n}^3 + 
             \left.\chi\right|_0 k_{p}^3\right)~.
\end{equation} 
In the above expressions, $c_{\sigma}^2 = (g_{\sigma}/m_{\sigma})^2$, $c_{\omega}^2=(g_\omega/m_\omega)^2$,  $c_{\rho}^2=(g_\rho/m_\rho)^2$ and
\begin{eqnarray}
\label{m star eqn}
    \left.m_*\right|_0 &=& m_*(k_n,k_p,0) \cr
         && \cr
        &=& m - \left.m_*\right|_0 \frac{c_\sigma^2}{2 \pi^2}  
            \left(k_n \sqrt{k_n^2 + \left.m^2_*\right|_0} + k_p 
            \sqrt{k_p^2 + \left.m^2_*\right|_0} + \frac{1}{2} 
            \left.m_*^2\right|_0 {\rm ln} \left[\frac{- k_n + 
            \sqrt{k_n^2 + \left.m^2_*\right|_0}}{k_n + 
            \sqrt{k_n^2 + \left.m^2_*\right|_0}}\right] \right. \cr
         && \cr
         && + \frac{1}{2} \left.\left.m_*^2\right|_0 {\rm ln} \left[\frac{- 
            k_p + \sqrt{k_p^2 + \left.m^2_*\right|_0}}{ k_p + 
            \sqrt{k_p^2 + \left.m^2_*\right|_0}}\right]\right) \ + b  m  c_\sigma^2\left(m - m_*\right)^2 +   c  c_\sigma^2\left(m - m_*\right)^3~. 
            \label{mstar}
\end{eqnarray}

The expressions for the other matter coefficients (see \cite{Kheto2014,Kheto2015}) that are used as the 
inputs in field equations are the following.
\begin{eqnarray}
{\cal A}|_0 &=& c_{\omega}^2-\frac{1}{4} c_{\rho}^2 + \frac{c^2_{\omega} }{ 5 
        \left.\mu^2\right|_0} \left(2 k_p^2 \frac{\sqrt{k_n^2 + 
        \left.m^2_*\right|_0}}{\sqrt{k_p^2 + 
        \left.m^2_*\right|_0}} + \frac{c^2_{\omega}}{3 \pi^2} 
        \left[\frac{k_n^2 k_p^3}{\sqrt{k_n^2 + 
        \left.m^2_*\right|_0}} + \frac{k_p^2 k_n^3}{\sqrt{k_p^2 + 
        \left.m^2_*\right|_0} }\right]\right)\cr
         && \cr
         &&
        +\frac{c^2_{\rho} }{ 20 
        \left.\mu^2\right|_0} \left(2 k_p^2 \frac{\sqrt{k_n^2 + 
        \left.m^2_*\right|_0} }{\sqrt{k_p^2 + 
        \left.m^2_*\right|_0}} + \frac{c^2_{\rho} }{ 12 \pi^2} 
        \left[\frac{k_n^2 k_p^3 }{ \sqrt{k_n^2 + 
        \left.m^2_*\right|_0}} + \frac{k_p^2 k_n^3 }{\sqrt{k_p^2 + 
        \left.m^2_*\right|_0} }\right]\right)\cr
         && \cr 
         &&
         -\frac{c^2_{\rho}c^2_{\omega} }{ 30\left.\mu^2\right|_0 \pi^2} 
        \left[\frac{k_n^2 k_p^3 }{ \sqrt{k_n^2 + 
        \left.m^2_*\right|_0}} - \frac{k_p^2 k_n^3 }{ \sqrt{k_p^2 + 
        \left.m^2_*\right|_0} }\right]  + \frac{3 \pi^2 k_p^2 
        }{ 5 \left.\mu^2\right|_0 k_n^3} \frac{k_n^2 + 
        \left.m^2_*\right|_0 }{ \sqrt{k_p^2 + 
        \left.m^2_*\right|_0}} \ , \\
        && \cr
{\cal B}|_0 &=& \frac{3 \pi^2 \left.\mu\right|_0 }{ k_n^3} - 
        c_{\omega}^2 \frac{k_p^3 }{ k_n^3}+ \frac{1}{4}c_{\rho}^2 \frac{k_p^3 }{ k_n^3} - \frac{c^2_{\omega} k_p^3 
        }{ 5 \left.\mu^2\right|_0 k_n^3} \left(2 k_p^2 
        \frac{\sqrt{k_n^2 + \left.m^2_*\right|_0} }{ \sqrt{k_p^2 + 
        \left.m^2_*\right|_0}} + \frac{c^2_{\omega} }{ 3 \pi^2} 
        \left[\frac{k_n^2 k_p^3}{ \sqrt{k_n^2 + 
        \left.m^2_*\right|_0}} + \frac{k_p^2 k_n^3 }{ \sqrt{k_p^2 + 
        \left.m^2_*\right|_0} }\right]\right)  \cr
        && \cr
        && - \frac{c^2_{\rho} k_p^3 
        }{ 20 \left.\mu^2\right|_0 k_n^3} \left(2 k_p^2 
        \frac{\sqrt{k_n^2 + \left.m^2_*\right|_0} }{ \sqrt{k_p^2 + 
        \left.m^2_*\right|_0}} + \frac{c^2_{\rho} }{ 12 \pi^2} 
        \left[\frac{k_n^2 k_p^3}{ \sqrt{k_n^2 + 
        \left.m^2_*\right|_0}} + \frac{k_p^2 k_n^3 }{\sqrt{k_p^2 + 
        \left.m^2_*\right|_0} }\right]\right)  \cr
        && \cr
        && + \frac{c^2_{\rho} c^2_{\omega}k^3_p }{ 30 \pi^2\left.\mu^2\right|_0 k_n^3} 
        \left[\frac{k_n^2 k_p^3 }{ \sqrt{k_n^2 + 
        \left.m^2_*\right|_0}} - \frac{k_p^2 k_n^3 }{ \sqrt{k_p^2 + 
        \left.m^2_*\right|_0} }\right]-\frac{3 \pi^2 k_p^5 }{ 5 \left.\mu^2\right|_0 k_n^6} 
        \frac{k_n^2 + \left.m^2_*\right|_0 }{ \sqrt{k_p^2 + 
        \left.m^2_*\right|_0}}\,\label{b00} \ , \\
        && \cr
{\cal C}|_0 &=& \frac{3 \pi^2 \left.\chi\right|_0 }{ k_p^3}+ \frac{1}{4}c_{\rho}^2 \frac{k_n^3 }{ k_p^3} - 
        c_{\omega}^2 \frac{k_n^3}{k_p^3} - \frac{c^2_{\omega} k_n^3 
        }{ 5 \left.\mu^2\right|_0 k_p^3} \left(2 k_p^2 
        \frac{\sqrt{k_n^2 + \left.m^2_*\right|_0} }{ \sqrt{k_p^2 + 
        \left.m^2_*\right|_0}} + \frac{c^2_{\omega} }{ 3 \pi^2} 
        \left[\frac{k_n^2 k_p^3 }{ \sqrt{k_n^2 + 
        \left.m^2_*\right|_0}} + \frac{k_p^2 k_n^3 }{ \sqrt{k_p^2 + 
        \left.m^2_*\right|_0} }\right]\right)  \cr
        && \cr
        && - \frac{c^2_{\rho} k_n^3 
        }{ 20 \left.\mu^2\right|_0 k_p^3} \left(2 k_p^2 
        \frac{\sqrt{k_n^2 + \left.m^2_*\right|_0} }{ \sqrt{k_p^2 + 
        \left.m^2_*\right|_0}} + \frac{c^2_{\rho} }{ 12 \pi^2} 
        \left[\frac{k_n^2 k_p^3}{ \sqrt{k_n^2 + 
        \left.m^2_*\right|_0}} + \frac{k_p^2 k_n^3 }{ \sqrt{k_p^2 + 
        \left.m^2_*\right|_0} }\right]\right)\cr
        && \cr
        && +\frac{c^2_{\rho} c^2_{\omega}k_n^3 }{ 30 \pi^2\left.\mu^2\right|_0 k_p^3} 
        \left[\frac{k_n^2 k_p^3 }{ \sqrt{k_n^2 + 
        \left.m^2_*\right|_0}} - \frac{k_p^2 k_n^3 }{ \sqrt{k_p^2 + 
        \left.m^2_*\right|_0} }\right]- \frac{3 \pi^2 }{ 5 \left.\mu^2\right|_0 k_p} 
        \frac{k_n^2 + \left.m^2_*\right|_0 }{ \sqrt{k_p^2 + 
        \left.m^2_*\right|_0}}~,
\label{coo} 
\end{eqnarray}

\begin{eqnarray}
{{\cal A}_0^0}|_0 &=-& \frac{\pi^4}{k_p^2k_n^2} \left.\frac{\partial^2 \Lambda}{\partial k_p\partial k_n}\right|_0
        = c_\omega^2 - \frac{c_\rho^2}{4}+ \frac{\pi^2 }{ k^2_p} \frac{ 
        \left.m_*\right|_0 \left.\frac{\partial m_* }{ \partial k_p}
        \right|_0 }{ \sqrt{k^2_n + \left.m^2_*\right|_0}}\ , \\
        && \cr
{{\cal B}_0^0}|_0 &=& \frac{\pi^4}{k_n^5} \left(\left.2\frac{\partial \Lambda}{\partial k_n}\right|_0-k_n\left.\frac{\partial^2 \Lambda}{\partial k_n^2}\right|_0\right) 
        = c_\omega^2 + \frac{c_\rho^2}{4} + \frac{\pi^2 }{ k^2_n} \frac{k_n + 
        \left.m_*\right|_0 \left.\frac{\partial m_* }{ \partial k_n}
        \right|_0 }{ \sqrt{k^2_n + \left.m^2_*\right|_0}} \ , \\
        && \cr
{{\cal C}_0^0}|_0 &=& \frac{\pi^4}{k_p^5} \left(\left.2\frac{\partial \Lambda}{\partial k_p}\right|_0-k_p\left.\frac{\partial^2 \Lambda}{\partial k_p^2}\right|_0\right)
        = c_\omega^2 +  \frac{c_\rho^2}{4} + \frac{\pi^2 }{ k^2_p} \frac{k_p + 
        \left.m_*\right|_0 \left.\frac{\partial m_* }{ \partial k_p}
        \right|_0 }{ \sqrt{k^2_p + \left.m^2_*\right|_0}} + 
        \frac{\pi^2 }{ k_p} \frac{1 }{ \sqrt{k^2_p + m^2_e}},
\end{eqnarray}
where,
\begin{eqnarray}
     \left.\frac{\partial m_* }{ \partial k_n}\right|_0 &=& - 
          \frac{c_\sigma^2 }{ \pi^2} \frac{\left.m_*\right|_0 k_n^2 
          }{ \sqrt{k_n^2 + \left.m^2_*\right|_0}} \left(\frac{3 m - 2 
          \left.m_*\right|_0 +3 b m c_\sigma^2\left(m-\left.m_*\right|_0\right)^2 
          +3 c  c_\sigma^2\left(m-\left.m_*\right|_0\right)^3}{\left.m_*\right|_0}\right.\nonumber \\ 
          &&\left.- \frac{c_\sigma^2 
          }{\pi^2} \left[\frac{k_n^3 }{ \sqrt{k_n^2 + \left.m^2_*
          \right|_0}} + \frac{k_p^3 }{ \sqrt{k_p^2 + 
          \left.m^2_*\right|_0}}\right]+2 b m c_\sigma^2\left(m-\left.m_*\right|_0\right) 
          +3 c  c_\sigma^2\left(m-\left.m_*\right|_0\right)^2\right)^{- 1}, 
          \\
 \textnormal{and,} \nonumber \\
 	&& \cr
     \left.\frac{\partial m_* }{ \partial k_p}\right|_0 &=& - 
          \frac{c_\sigma^2 }{ \pi^2} \frac{\left.m_*\right|_0 k_p^2 
          }{ \sqrt{k_p^2 + \left.m^2_*\right|_0}} \left(\frac{3 m - 2 
          \left.m_*\right|_0 +3 b m c_\sigma^2\left(m-\left.m_*\right|_0\right)^2 
          +3 c  c_\sigma^2\left(m-\left.m_*\right|_0\right)^3}{\left.m_*\right|_0}\right.\nonumber \\ 
          &&\left.- \frac{c_\sigma^2 
          }{\pi^2} \left[\frac{k_n^3 }{ \sqrt{k_n^2 + \left.m^2_*
          \right|_0}} + \frac{k_p^3 }{ \sqrt{k_p^2 + 
          \left.m^2_*\right|_0}}\right]+2 b m c_\sigma^2\left(m-\left.m_*\right|_0\right) 
          +3 c  c_\sigma^2\left(m-\left.m_*\right|_0\right)^2\right)^{- 1}, 
\end{eqnarray}
respectively.


\begin{thebibliography}{99}
\bibitem{Abbott2017} B. P. Abbott {\it et al.} (LIGO Scientific and Virgo Collaborations), Phys. Rev. Lett. {\bf 119}, 161101 (2017).
\bibitem{Agathos2015} M.  Agathos,  J.  Meidam,  W.  Del  Pozzo,  T.  G.  F.  Li, M. Tompitak, J. Veitch, S. Vitale,  and C. Van Den Broeck, Phys. Rev. D {\bf 92} , 023012 (2015).
\bibitem{Takami2014} K. Takami, L. Rezzolla, and L. Baiotti, Phys. Rev. Lett. {\bf 113}, 091104 (2014)
\bibitem{Bose2018}S. Bose, K. Chakravarti, L. Rezzolla, B. S. Sathyaprakash, K. Takami, Phys. Rev. Lett. {\bf 120}, 031102 (2018)
\bibitem{Flanagan2008}\'{E}. \'{E} Flanagan, and T. Hinderer, Phys. Rev. D {\bf 77}, 021502 (2008).
\bibitem{Hinderer2008} T. Hinderer, Astrophys. J. {\bf 677}, 1216 (2008).
\bibitem{Binnington2009} T. Binnington, E. Poisson Phys. Rev. D {\bf 80}, 084018 (2009)
\bibitem{Damour2009} T. Damour, A. Nagar, Phys. Rev. D {\bf80}, 084035 (2009)
\bibitem[Hinderer et al. (2010)]{Hinderer2010} T. Hinderer, B. D. Lackey, R. N. Lang and J. S. Read, Phys. Rev. D {\bf 81}, 123016 (2010).
\bibitem{Vines2011} J. Vines, E. E. Flanagan, and T. Hinderer, Phys. Rev. D {\bf 83}, 084051 (2011)
\bibitem{Damour2012} T. Damour, A. Nagar, and L. Villain, Phys. Rev. D {\bf 85}, 123007 (2012) 
\bibitem{Read2013} J. S. Read, L. Baiotti, J. D. E. Creighton, J. L. Friedman, B. Giacomazzo, K. Kyutoku, C. Markakis, L. Rezzolla, M. Shibata, and K. Taniguchi, Phys. Rev. D {\bf 88}, 044042 (2013).
\bibitem{DelPozzo2013}W. Del Pozzo, T. G. F. Li, M. Agathos, C. Van Den Broeck, and S. Vitale, Phys. Rev. Lett. {\bf 111}, 071101 (2013).
\bibitem{Wade2014}L. Wade, J. D. E. Creighton, E. Ochsner, B. D. Lackey, B. F. Farr, T. B. Littenberg, and V. Raymond, Phys. Rev. D {\bf 89} , 103012 (2014)
\bibitem{Favata2014} M. Favata, Phys. Rev. Lett. {\bf 112}, 101101 (2014).
\bibitem{Hotokezaka2016} K. Hotokezaka, K. Kyutoku, Y. Sekiguchi, and M. Shibata, Phys. Rev. D {\bf 93}, 064082 (2016)
\bibitem{Migdal1959}A. B. Migdal, Nucl. Phys. {\bf 13}, 655 (1959); 
\bibitem{Clark1992}J. Clark, R. Dav\'e, and J. Chen, The Structure and Evolution of Neutron Stars, edited by D. Pines, R. Tamagaki, and S. Tsurate (Addison-Wesley, Redwood City, CA, 1992).
\bibitem{Sedrakian2018} A. Sedrakian and J. W. Clark, arXiv:1802.00017v1 [nucl:th]
\bibitem{Baym1975} G. Baym, C. J. Pethick, D. Pines, and M. Ruderman, Nature (London) {\bf 224}, 872 (1969).
\bibitem{Anderson1975} P. W. Anderson and N. Itoh, Nature (London) {\bf 256}, 25 (1975)
\bibitem{Page2011} D. Page, M. Prakash, J. M. Lattimer, and A. W. Steiner, Phys. Rev. Lett. {\bf 106}, 081101 (2011)
\bibitem{Shternin2011} P. S. Shternin, D. G. Yakovlev, C. O. Heinke, W. C. G. Ho, and D. J. Patnaude, Mon. Not. R. Astron. Soc. {\bf 412}, L108 (2011);
\bibitem{Carter1989} B. Carter, Relativistic Fluid Dynamics, edited by A. Anile and M. Choquet-Bruhat ~Springer-Verlag, Berlin, 1989.
\bibitem{Comer1994} G.L. Comer and D. Langlois, Class. Quantum Grav. {\bf 11}, 709 (1994).
\bibitem{Carter1995}  B. Carter and D. Langlois, Phys. Rev. D {\bf 51}, 5855 (1995).
\bibitem{carter1998 1} B. Carter and D. Langlois, Nucl. Phys. {\bf B454}, 402 (1998).
\bibitem{carter1998 2} B. Carter and D. Langlois, Nucl. Phys. {\bf B531}, 478 (1998).
\bibitem{Langois1998} D. Langlois, A. Sedrakian, and B. Carter, Mon. Not. R. Astron. Soc. {\bf 297}, 1189 1998.
\bibitem{Prix2000} R. Prix, Phys. Rev. D {\bf 62}, 103005 (2000).
\bibitem{Sauls1989}J. Sauls,  Timing Neutron Stars, edited by H. \"Ogelman and E. P. J. van den Heuvel (Kluwer, Dordrecht, 1989)
\bibitem{Andersson2001} N. Andersson and G.L. Comer, Class. Quantum Grav. {\bf 18}, 969 (2001).
\bibitem{Prix2002} R. Prix, G.L. Comer, and N. Andersson, Astron. Astrophys. {\bf 381}, 178 (2002).
\bibitem{Comer1999} G. Comer, D. Langlois and L. M. Lin, Phys. Rev.  D {\bf 60}, 104025 (1999).
\bibitem{Comer2002} G. L. Comer, Found. Phys. {\bf 32}, 1903 (2002).
\bibitem{Andersson2002} N. Andersson, G. L. Comer, and D. Langois, Phys. Rev. D {\bf 66}, 104002 (2002).
\bibitem{Yu2017_1} H. Yu, and N. N. Weinberg, Mon. Not. R. Astron. Soc. {\bf 464}, 2622 (2017)
\bibitem{Yu2017_2} H. Yu, and N. N. Weinberg, Mon. Not. R. Astron. Soc. {\bf 470}, 350 (2017)
\bibitem{Comer2003} G. L. Comer and R. Joynt, Phy. Rev. D {\bf 68}, 023002 (2003).
\bibitem{Comer2004} G. Comer,  Phys. Rev. D {\bf 69}, 123009 (2004).
\bibitem{Kheto2014} A. Kheto and D. Bandyopadhyay, Phys. Rev. D {\bf 89}, 023007 (2014).
\bibitem{Kheto2015} A. Kheto and D. Bandyopadhyay, Phys. Rev. D {\bf 91}, 043006 (2015).
\bibitem{Thorne1967} K. S. Thorne and A. Campolattaro, Astrophys. J. {\bf 149}, 591 (1967)
\bibitem{Regge1957}T. Regge and J. A. Wheeler, Phys. Rev. {\bf 108}, 1063 (1957).
\bibitem{Chandra} S. Chandrasekhar, \textit{The Mathematical Theory of Black Holes}~(Oxford University press, New Delhi 2010)
\bibitem{Fattoyev2010} F.J. Fattoyev, C.J. Horowitz, J. Piekarewicz, and G. Shen, Phys. Rev. C {\bf 82}, 055803 (2010).
\bibitem{Glendenning1991}N. K. Glendenning and S. A. Moszkowski, Phys. Rev. Lett. {\bf 67}, 2414 (1991).


\end{thebibliography}
\end{document}